\documentclass[11pt,letterpaper]{article}

\usepackage{graphicx,array}
\usepackage{url}
\usepackage{color}
\usepackage{latexsym}
\usepackage{amsthm}
\usepackage{amsmath}
\usepackage{amssymb}
\usepackage{amsfonts}
\usepackage[numbers,sort&compress]{natbib}
\usepackage{bm}
\usepackage{slashed}
\usepackage{mathrsfs}
\usepackage{enumerate}
\usepackage{tikz}
\usepackage{siunitx}
\usepackage{mdframed}
\usepackage{setspace}  
\usepackage{esvect}
\usepackage{esint}
\usepackage{subcaption}
\usepackage[normalem]{ulem}
\usepackage{setspace}

\usepackage{soul}

\usepackage[utf8x]{inputenc}%
\usepackage{tcolorbox}%
\usepackage[normalem]{ulem} 

\usepackage{hyperref} 
\hypersetup{
    colorlinks=true,       
    linkcolor=red,          
    citecolor=blue,        
    filecolor=magenta,      
    urlcolor=blue           
}
\usepackage[all]{hypcap} 

\usepackage{natbib}
\newcommand{\nc}{\newcommand}

\nc{\beq}{\begin{equation}}  
\nc{\eeq}{\end{equation}}  
\nc{\beqa}{\begin{eqnarray}}  
\nc{\eeqa}{\end{eqnarray}}  
\nc{\bit}{\begin{itemize}}  
\nc{\eit}{\end{itemize}}  

\newcommand{\eg}{{\it e.g.}}
\newcommand{\ie}{{\it i.e.}}

\usepackage{floatrow}
\newfloatcommand{capbtabbox}{table}[][\FBwidth]

\usepackage{blindtext}

\def\figureautorefname~#1\null{Fig.\,#1\null}
\def\tableautorefname~#1\null{Tab.\,#1\null}

\def\equationautorefname~#1\null{Eq.\,(#1)\null}

\title{ 
{\bf Cosmological Constraints on First-Order \\Phase Transitions}
\author{\large Yang Bai and Mrunal Korwar}
\date{\small \it 
Department of Physics, University of Wisconsin-Madison, Madison, WI 53706, USA}
}

\begin{document}

\maketitle

\setlength{\parskip}{0.2ex}

\begin{abstract}	
First-order phase transitions exist in many models beyond the Standard Model and can generate detectable stochastic gravitational waves for a strong one. Using the cosmological observables in big bang nucleosynthesis and cosmic microwave background, we derive constraints on the phase transition temperature and strength parameter in a model-independent way. For a strong phase transition, we find that the phase transition temperature should be above around 2 MeV for both reheating photon and neutrino cases. For a weak one with the temperature below 1 MeV, the phase transition strength parameter is constrained to be smaller than around 0.1. Implications for using a first-order phase transition to explain the NANOGrav observed gravitational wave signal are also discussed. 
\end{abstract}

\thispagestyle{empty}  
\newpage  
  
\setcounter{page}{1}  


\section{Introduction}\label{sec:Introduction}

Phase transition in the early universe is an important subject to answer many unsolved questions in nature including the origin of the baryon asymmetry and dark matter. Within the Standard Model (SM), both the electroweak and QCD phase transitions are cross-over~\cite{Dine:1992wr,Arnold:1992rz,Fodor:2001pe}. Beyond the SM (BSM), first-order phase transitions (FOPTs) are ubiquitous including in models with extension of the Higgs sector for electroweak baryogenesis~\cite{Cohen:1993nk,Bodeker:2020ghk} and dark QCD-like confinement to explain dark matter properties~\cite{Bai:2013xga,DeGrand:2015zxa,Kribs:2016cew} (\ie, changing dark matter abundance from bubble dynamics~\cite{Bai:2018dxf,Hong:2020est,Asadi:2021pwo,Gross:2021qgx}). Other than leaving imprints in dark matter, FOPT can also generate observable stochastic gravitational waves if it is a strong one and affect big bang nucleosynthesis (BBN) or cosmic microwave background (CMB) physics if the phase transition temperature is low. 

Measuring the stochastic gravitational wave background (SGWB) has been demonstrated as a promising way to probe early universe FOPT~\cite{Witten:1984rs} (see Refs.~\cite{Schwaller:2015tja,Tsumura:2017knk,Aoki:2017aws,Croon:2018erz,Helmboldt:2019pan,Croon:2020cgk,Bigazzi:2020avc} for recent studies). With the first direct measurement of gravitational waves by the LIGO-Virgo collaboration in 2016~\cite{LIGOScientific:2016aoc}, a new era of gravitational wave astronomy has opened up.  The current ground-based and future space-based gravitational wave experiments cover many orders of magnitude in frequency from $10^{-9}$~Hz to $10^3$~Hz and could probe the phase transition temperatures from keV to PeV scales. For the low frequency or temperature region, it is specially interesting because of many existing and future pulsar timing array experiments (PTAs) including NANOGrav~\cite{McLaughlin:2013ira, Brazier:2019mmu}, EPTA~\cite{Lentati:2015qwp}, PPTA~\cite{Hobbs:2013aka}, MeerTime~\cite{Bailes:2018azh} and CHIME~\cite{Ng:2017djg}. Recently, the NANOGrav collaboration has observed a signal of SGWB in their 12.5 year dataset~\cite{NANOGrav:2020bcs}, which could be explained by the tightly-bound inspiralling supermassive binary black holes (SMBHBs) as well as the common sources from a strong first-order cosmological phase transition with a low phase transition temperature as low as the MeV scale~\cite{Ratzinger:2020koh,Li:2021qer,Bian:2020urb}. 

Cosmological observables in BBN and CMB physics can also probe phase transitions with a low temperature. For a strong FOPT, the universe could be temporarily in a vacuum energy dominated one and has the background Hubble evolution different from the ordinary cosmology. If this happens very late and close to the BBN time, the primordial light element abundance will be altered from the standard cosmology prediction. Another and maybe more important effect is that the vacuum energy contained in the false vacuum will be eventually converted into radiation energy to reheat either photons or neutrinos (it can also reheat dark radiation, which leads to a simpler constraint and will also be discussed here). If this happens around or after photon and neutrino thermally decoupling time of around one second, either photon or neutrino temperature will be different from the standard cosmology predictions. The effective radiation degrees of freedom, $N_{\rm eff}$, can then be modified, which subsequently changes the light element abundances and the power spectra that are measured by the CMB experiments. 

In this article, we use the cosmological observables to constrain a wide range of models with a FOPT. For a strong FOPT and model-independently, we will demonstrate that the phase transition temperature is constrained to be above around $2$~MeV based on the current cosmological data. Our study is the first one to serve this purpose and has the immediate consequence of restricting the parameter space to use a phase transition to explain the NANOGrav observed SGWB. This study is another example demonstrating that the cosmological data is superb on probing various BSM physics (\eg, low-temperature inflationary reheating~\cite{deSalas:2015glj, Ichikawa:2005vw, Kawasaki:2000en} and MeV-scale thermal dark sectors~\cite{Breitbach:2018ddu, Sabti_2020, Giovanetti:2021izc}).

\section{First-order phase transition}
\label{sec:phase:transition}
There are many models providing a first-order phase transition in the hidden or visible sector. 
In this work, we will derive the constraints from BBN and CMB observables for a class of models and want to keep the presentation as model-independent as possible. In our later discussion, we will use the visible sector temperature to keep tracking the cosmological history. For convenience, we define two relevant temperatures: a) $T_{\gamma}^{\rm p}$ is the percolation temperature (of photon) when 34\% of the volume in the universe is converted to the true minimum; b) $T^{\rm rh}_{\gamma}$ and $T^{\rm rh}_{\nu}$ are the reheating temperatures for photon and neutrino after the energy contained by the order-parameter field is converted into photons (including charged-leptons and nucleons) or neutrinos. For simplicity and also to derive a conservative bound, we will work on the instantaneous reheating scenarios with a negligible time interval between the percolation time and the reheating time. We also do not consider the photon and neutrino comparable-weighted reheating case, which is unlikely based on different particle physics models.

To quantify the strength of the phase transition, we introduce a strength parameter $\alpha_*$ that is defined as the ratio of the vacuum over the relativistic energy density at the percolation time or $\alpha_* \approx \Delta V(T_{\gamma}^{\rm p})/\rho_{\rm R}(T_{\gamma}^{\rm p})$ for $\alpha_* \gtrsim 1$~\cite{Ellis:2018mja}. Here, $\Delta V(T)$ is the effective potential difference between the two vacua and $\rho_{\rm R}(T) = (\pi^2/30)g^{\rm t}_*\,T_\gamma^4$ is the total radiation energy density. $g^{\rm t}_*$ is the total effective radiation degrees of freedom and should take into account of possible different temperatures for neutrinos and/or hidden-sector radiation particles.

After the phase transition or the percolation time, the vacuum energy can be converted into several possible radiation energies. The first possibility is that some hidden-sector radiation degrees of freedom are heated up. If they are stable, they will contribute significantly to $\Delta N_{\rm eff}(t_{\rm CMB})$ at the CMB period with 
\beqa
\Delta N_{\rm eff}(t_{\rm CMB}) 
&=&  \frac{8}{7} \left(\frac{11}{4}\right)^{4/3}\, \frac{g^{4/3}_{*s,\gamma}(t_{\rm CMB})}{2\,g^{1/3}_{*,\gamma}(t_{\rm rh})}
\left[ \alpha_* + (1+\alpha_*) \frac{g_{*,h}(t_{\rm rh})\,(T_{h}^{\rm p})^4}{g_{*,\gamma}(t_{\rm rh})\,(T_{\gamma}^{\rm p})^4} \right] ~.
\eeqa
Here, $g_{*,h}$ is the radiation degrees of freedom in the hidden sector and $T_{h}^{\rm p}$ is the hidden sector temperature at the percolation time. In the extremal case with a chilly hidden sector with $T^{\rm p}_h \ll T^{\rm p}_\gamma$, one has $\Delta N_{\rm eff}(t_{\rm CMB}) > 6.3\,(2.9)\times\alpha_*$ for $g_{*s,\gamma(t_{\rm CMB})} \approx 4$ and $g_{*,\gamma(t_{\rm rh})} \approx 10.75\,(106.75)$ for $T^{\rm p}_\gamma$ of a few MeV\,(above the top quark mass). Applying the limit  from the CMB epoch given by the Planck 2018 observations: $\Delta N_{\text{eff}}< 0.51$~\cite{Aghanim:2018eyx,Riess_2018}, one has $\alpha_* < 0.08\,(0.18)$ for the two choices of reheating time. 

Given the stringent constraints on the case of reheating the hidden-sector radiation, we switch our focus to the case with reheating the SM particles. For a low reheating temperature $\mathcal{O}(\mbox{MeV})$, both possibilities of reheating photons (or other SM particles in the same plasma as photon) and reheating neutrinos are to be considered (see Ref.~\cite{Hannestad:2004px} for constraints on the inflaton reheating temperature). Leaving the particle physics realization aside and for the photon-reheating case, one has $T^{\rm rh}_{\nu} = T^{\rm p}_{\nu}$ and 
\beqa\label{eq:phtotonreh}
&&T^{\rm rh}_{\gamma} = \left[ 1 + \alpha_* + \alpha_* \dfrac{g_{*,\nu}(t_{\rm rh})}{g_{*,\gamma}(t_{\rm rh})} \left( \dfrac{T^{\rm p}_{\nu}}{T^{\rm p}_{\gamma}}\right)^4  \right]^{1/4}\, T^{\rm p}_{\gamma} \,, 
\qquad\qquad \mbox{[photon reheating]}~. 
\eeqa
Here, $g_{*,\nu}(t_{\rm rh})\approx 21/4$ and $g_{*,\gamma}(t_{\rm rh})\approx 11/2$ for both temperatures above MeV. For this case and after reheating, the neutrino temperature is relatively lower than the photon one. For the other case with reheating neutrinos, one has $T^{\rm rh}_{\gamma} = T^{\rm p}_{\gamma}$ and 
\beqa\label{eq:neutrinoreh}
&&T^{\rm rh}_{\nu} = \left[ 1 + \alpha_* + \alpha_* \dfrac{g_{*,\gamma}(t_{\rm rh})}{g_{*,\nu}(t_{\rm rh})} \left( \dfrac{T^{\rm p}_\gamma}{T^{\rm p}_\nu}\right)^4  \right]^{1/4}\, T^{\rm p}_\nu \,, 
\qquad \qquad \mbox{[neutrino reheating]}~,
\eeqa
which could lead to a higher neutrino temperature than the photon one. Here, we do not distinguish different flavors of neutrinos because the neutrino oscillations and interactions (the SM or BSM one) thermalize the neutrino sector.

\section{{\boldmath$N_{\rm eff}$} and BBN observables}
\label{sec:BBN}

The time dependence of photon and neutrino temperatures can be calculated based on their Friedmann equations allowing the energy transfer between them (see Appendix~\ref{sec:Temp_evol} for detailed formulas and calculations). Using the asymptotic photon and neutrino temperatures at a later time, one can calculate the effective number of relativistic degrees of freedom
\beqa
N_{\rm eff} = \frac{8}{7} \left(\frac{11}{4}\right)^{4/3}\frac{\rho_{\nu}}{\rho_{\gamma}}=3\, \left(\frac{11}{4}\right)^{4/3}\,\times\,\frac{T^4_\nu(t_{\rm today})}{T^4_\gamma(t_{\rm today})} \, .  
\eeqa
For the SM case, we have checked that our numerical calculations reproduce the SM value of $N_{\rm eff}=3.045$~\cite{Abenza_2020}. For the photon-reheating case, a smaller $T_{\gamma}^{\rm rh}$ (a later phase transition after the photon and neutrino thermally decouple time) or a larger $\alpha_*$ corresponds to a smaller $N_{\rm eff}$ compared to the SM case (see the left panel of Fig.~\ref{fig:PandNReh_Neff} in Appendix~\ref{sec:plots-observables}). This is because the ratio of the final neutrino over photon temperature is reduced more for a later phase transition. For the neutrino reheating case, an opposite behavior is observed with an increased $N_{\rm eff}$ as $T_{\nu}^{\rm rh}$ decreases or $\alpha_*$ increases (see the right panel of Fig.~\ref{fig:PandNReh_Neff} in Appendix~\ref{sec:plots-observables}). 

The modification of the background evolution due to a phase transition also changes the temperature and time relation and affects the light-element abundance during the BBN period. To calculate the modified predictions, we use the publicly available \texttt{PRIMAT} code~\cite{Pitrou_2018} to incorporate various weak and nuclear reaction rates with the calculated $T_{\gamma}(t)$ and $T_{\nu}(t)$ (as well as $a(t)$ based on the Hubble equation). As a sanity check and for the SM case, we use the central value of $\Omega_{b}h^{2}= 0.02230 \pm 0.00020$~\cite{Aghanim:2018eyx} and the neutron lifetime $\tau_{n} = 879.4 \pm 0.6$\,sec~\cite{Czarnecki:2018okw} to obtain $Y_{\rm P} \equiv \rho(^4\mbox{He})/\rho_b= 0.24703$ and the primary deuterium aboudnance $\rm D/ \rm H|_{P} = 2.463 \times 10^{-5}$, which agrees well with the observed value $Y_{\rm P}^{\rm obs} = 0.245 \pm 0.003$~\cite{ParticleDataGroup:2020ssz} and has a small ($\sim$\,$2\,\sigma$) tension with $\rm D/\rm H|_{P}^{\rm obs} = (2.547 \pm 0.025) \times 10^{-5}$~\cite{ParticleDataGroup:2020ssz} (this tension has been recently discussed in Ref.~\cite{Pitrou:2020etk}).

\begin{figure}[tb!]
	\centering
	\includegraphics[width=0.48\textwidth]{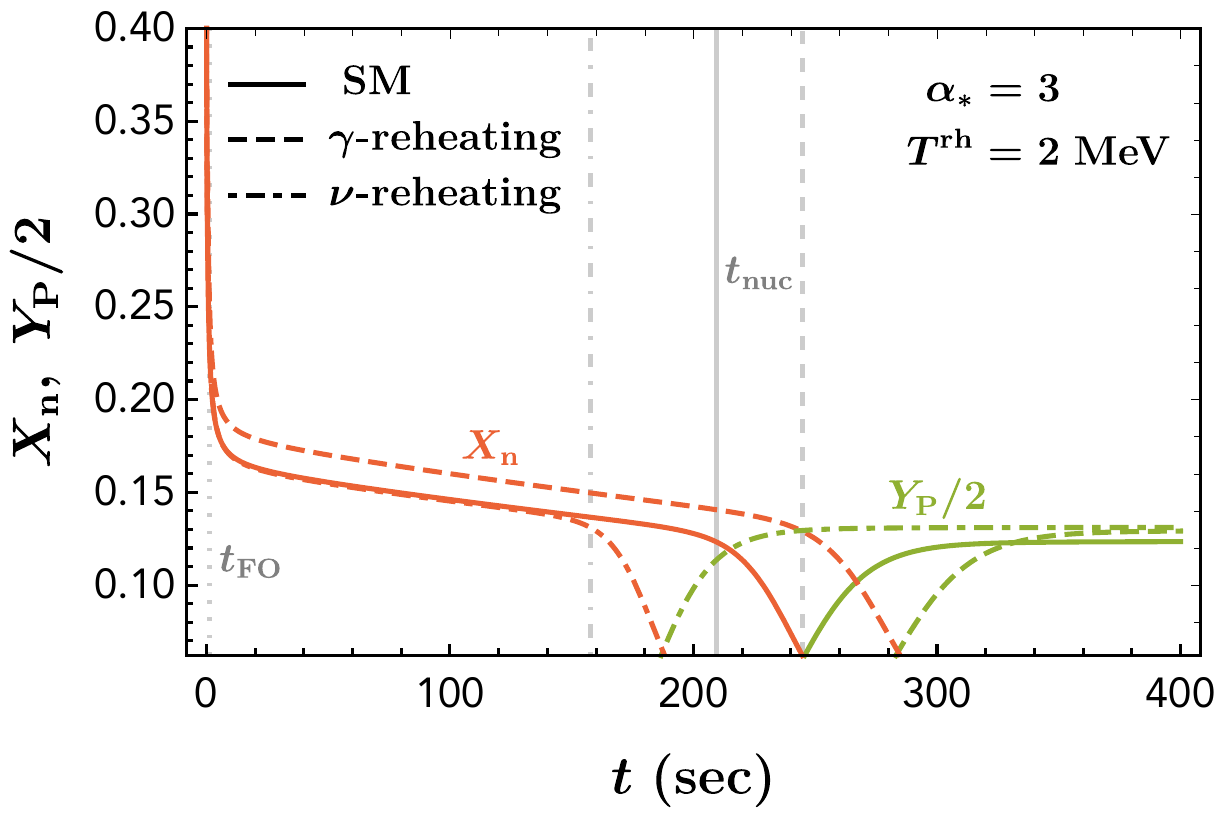} \hspace{3mm}
	\includegraphics[width=0.475\textwidth]{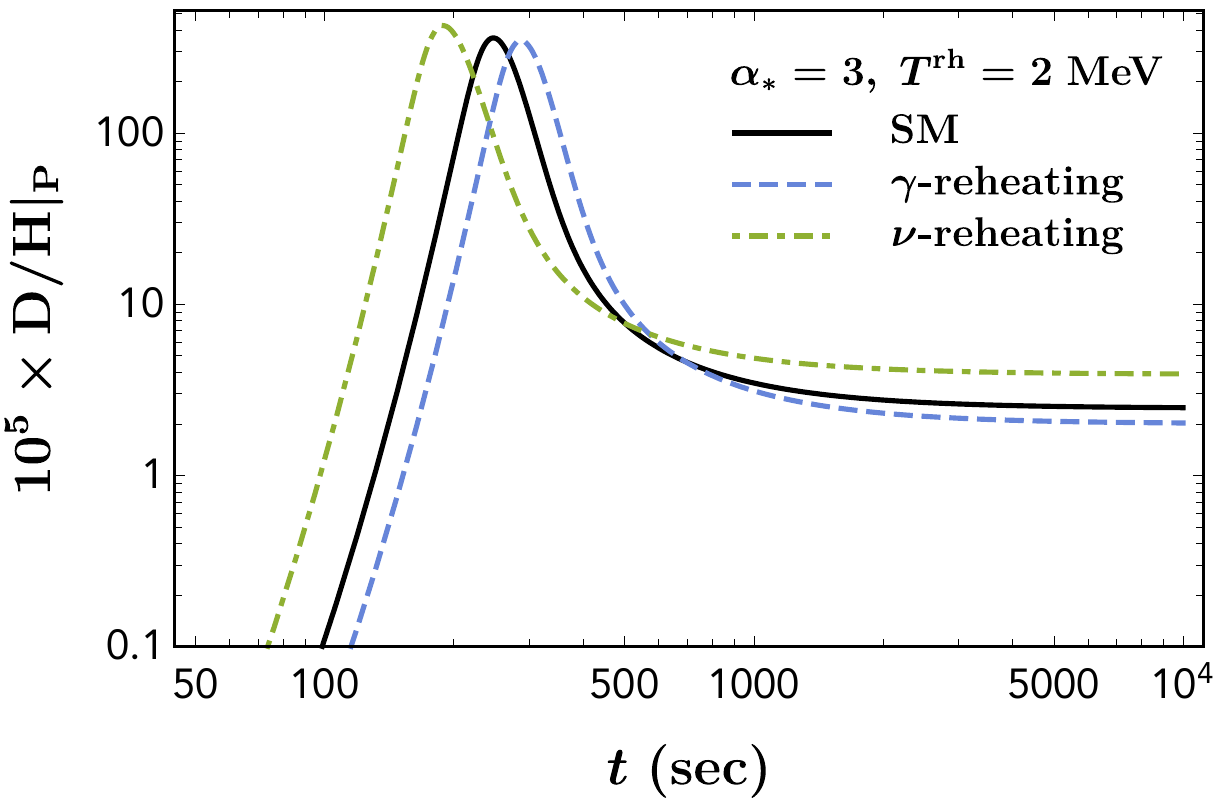}
		\caption{Left panel: $X_{n}$ and $Y_{\rm P}$ as a function of time for different models. The leftmost gray vertical line shows the neutron freeze-out time $t_{\rm FO}\sim 1$~sec, while the right three gray vertical lines are the nucleosynthesis starting time with $T_{\gamma}^{\rm nuc}\approx 0.078\,\mbox{MeV}$ for the three different models. 
Right panel: the deuterium abundance ratio $\rm D/ \rm H|_{P}$ as a function of time. In both panels, $\Omega_{b}h^{2}=0.0223$. }
	\label{fig:abundance}
\end{figure}

One of the important intermediate variables for light-element abundance is the neutron fraction $X_n \equiv n_n/n_b$, which provides an overall BBN timescale. In the left panel of Fig.~\ref{fig:abundance}, we show $X_n$ as a function of time for two representative phase transition cases as well as the SM case. For both cases, $X_n$ follows a similar thermal equilibrium distribution as in the SM for $T \gtrsim Q$ with $Q = 1.293$\,MeV, the neutron-proton mass difference. On the other hand, they have different overall magnitudes because the photon-reheating case has a higher photon (or baryon) plasma temperature and a higher $X_n$. 

As temperature drops,  $X_n$ has a freeze-out temperature, $T_{\gamma}^{\rm FO}$ around 0.8~MeV~\cite{Pitrou_2018} or $t_{\rm FO} \sim 1$~sec, when the weak interaction rate equals the Hubble rate. For the photon-reheating case, the neutrino temperature is reduced compare to the SM case. This means a smaller (electron) neutrino density and a smaller weak rate $\Gamma_{np}$. On the other hand, the Hubble rate is also reduced because of a reduced $N_{\rm eff}$. Those two effects partly cancel each other~\cite{Ichikawa:2005vw}. Because of the higher power dependence on $T$ for $\Gamma_{np}$, the modification on $\Gamma_{np}$ is more significant~\cite{Huang:2021dba}. As a result, the freeze-out temperature $T_{\gamma}^{\rm FO}$ is higher than the SM case. For the neutrino-reheating case, the changes for various rates are opposite to the photon case and $T_{\gamma}^{\rm FO}$ is smaller than the SM case. Earlier (later) freeze-out tends to have a larger (smaller) the neutron fraction at the freeze-out temperature $X_{n}(t_{\rm FO})$.

After freeze-out, $X_{n}$ continues to drop because of neutron decay with $X_{n}(t>t_{\rm FO})\approx X_{n}(t_{\rm FO})e^{-t/\tau_{n}}$ (the region between $t_{\rm FO}$ to $t_{\rm nuc}$ in the left panel of Fig.~\ref{fig:abundance}). When the temperature drops to the nucleosynthesis temperature $T_{\gamma}^{\rm nuc}\approx 0.078\,\mbox{MeV}$ (at $t_{\rm nuc}$), the fraction $Y_{\rm P}$ starts to grow (helium starts to be produced). The final $Y_{\rm p}$ abundance is determined by $X_{n}(t_{\rm nuc})$ at the starting time of nucleosynthesis $t_{\rm nuc}$ with the corresponding temperature $T_{\gamma}^{\rm nuc}$, or $Y_{\rm P} \approx 2X_{n}(t_{\rm nuc})$. For the photon-reheating case with a smaller $N_{\rm eff}$ and a smaller Hubble rate, one has a later $t_{\rm nuc}$, which prefers to have a smaller final $Y_{\rm p}$. However, this is over compensated by the larger $X_{n}(t_{\rm FO})$ and leads to a larger final $Y_{\rm p}$ than the SM case (see the end-time value of the dashed line in the left panel of Fig.~\ref{fig:abundance}). For the neutrino-reheating case with a larger $N_{\rm eff}$ and a larger Hubble rate, the nuclear reactions happen earlier with an earlier $t_{\rm nuc}$. So, the final $Y_{\rm p}$ is also larger than the SM case (see the upper two panels of Fig.~\ref{fig:PandNReh} in Appendix~\ref{sec:plots-observables} for $Y_{\rm p}$ with different $T^{\rm rh}$ and $\alpha_*$).

In the right panel of Fig.~\ref{fig:abundance}, we show the deuterium abundance ratio $\rm D/ \rm H|_{P}$ as a function of time for both photon- and neutrino-reheating cases as well as the SM case. The ``deuterium bottleneck" is clear from the peak structure of all three curves. The deuterium abundance ratio reaches the peak at around $T_{\gamma}^{\rm nuc}\approx 0.078\,\mbox{MeV}$ for all three cases (at this temperature, the deuterium equilibrium abundance from the Saha's formula reaches the peaked one~\cite{Mukhanov:2003xs}). Because of different temperature and time relations for the three cases, the peaks are located at different times. The photon-reheating case has a smaller $N_{\rm eff}$, a slower Hubble rate and a slower drop of the photon temperature, the peak arrives at a later time compared to the SM case. It also has a smaller late-time deuterium abundance because of a smaller $N_{\rm eff}$~\cite{Mukhanov:2003xs}. The neutrino-reheating case has an opposite behavior compared to the photon-reheating case (see the lower two panels of Fig.~\ref{fig:PandNReh} of Appendix~\ref{sec:plots-observables} for the deuterium abundance ratios for different model parameters). Comparing the modifications for $Y_{\rm P}$ and ${\rm D}/{\rm H}|_{\rm P}$, the deuterium abundance is likely to provide the dominate BBN constraint. Also, note that the deuterium abundance is approximately proportional to $N_{\rm eff}$. On the other hand, the helium abundance ratio does not have such a correlation with $N_{\rm eff}$ and hence provides additional constraints from $N_{\rm eff}$. 

\section{Cosmological constraints on phase transitions}
\label{sec:constraint}

In this section, we derive various constraints on the two phase transition model parameters: $\alpha_{*}$ and $T_{\gamma}^{\rm rh}$ (or $T_{\nu}^{\rm rh}$) using different sets of cosmological data. Since some observables also depend on the baryon abundance or the ordinary $\Omega_b h^2$ parameter, we will marginalize over $\Omega_b h^2$ to derive the constraints on the new physics model parameters. Specifically, we will choose the range $\Omega_b h^2 \in [0.0215,0.0230]$, which is approximately 3 sigma around the Planck measured value~\cite{Aghanim:2018eyx}. 

\vspace{-2mm}
\begin{itemize}
\item \textbf{BBN:}\; We use the PDG recommended values for $Y_{\rm P}$ and $\rm D/\rm H|_{\rm P}$ and their $1\sigma$ error bars~\cite{ParticleDataGroup:2020ssz}. The theory error is taken from Ref.~\cite{Pitrou:2020etk}  [see \eqref{eq:BBN-data} of Appendix~\ref{App:dataset} for detailed numbers]. We ignore the small correlation between the theory errors~\cite{Giovanetti:2021izc} and add the experimental and theoretical errors in quadrature. 
\vspace{-1mm}
    
 \item  \textbf{CMB and local $\bm{H_0}$:}\; Because of the disagreement between local~\cite{Riess:2019cxk} and CMB determination of the Hubble constant~\cite{Aghanim:2018eyx}, we consider two sets of dataset in our analysis (i) Planck only: using 2018 Planck baseline TTTEEE+lowE analysis~\cite{Aghanim:2018eyx,Planck:2019nip} [see \eqref{eq:planck-only} of Appendix~\ref{App:dataset}], and (ii) Planck + $H_0$: combining the Planck CMB data with Baryon Acoustic Oscillation (BAO) measurements~\cite{Beutler_2011,Ross_2015,BOSS:2016wmc} and local measurement of $H_0$ from the SH0ES collaboration~\cite{Riess:2019cxk} [see \eqref{eq:planck-H0} of Appendix~\ref{App:dataset}]. The numerical values for the three observables $\Theta \equiv (\Omega_{b}h^{2}, N_{\text{eff}}, Y_{\rm P})$ as well as their covariance matrix are taken from Ref.~\cite{Sabti_2020}.
 \vspace{-1mm}

 \item \textbf{CMB + BBN (+ $\bm{H_0}$):}\; the total $\chi^2$ is simply the summed one from CMB and BBN. 
 \vspace{-1mm}

\item \textbf{Future CMB experiments:}\; both the Simons Observatory~\cite{SimonsObservatory:2018koc} and CMB-S4~\cite{CMB-S4:2016ple, Abazajian:2019eic} are considered as examples for future CMB experiments. The fiducial mean values for the observables $\Theta$ as well as the covariance matrix can be found in Ref.~\cite{Sabti_2020} and also shown in Appendix~\ref{App:dataset}. 
\end{itemize}

\begin{figure}[tb!]
	\centering
\includegraphics[width=.48\textwidth]{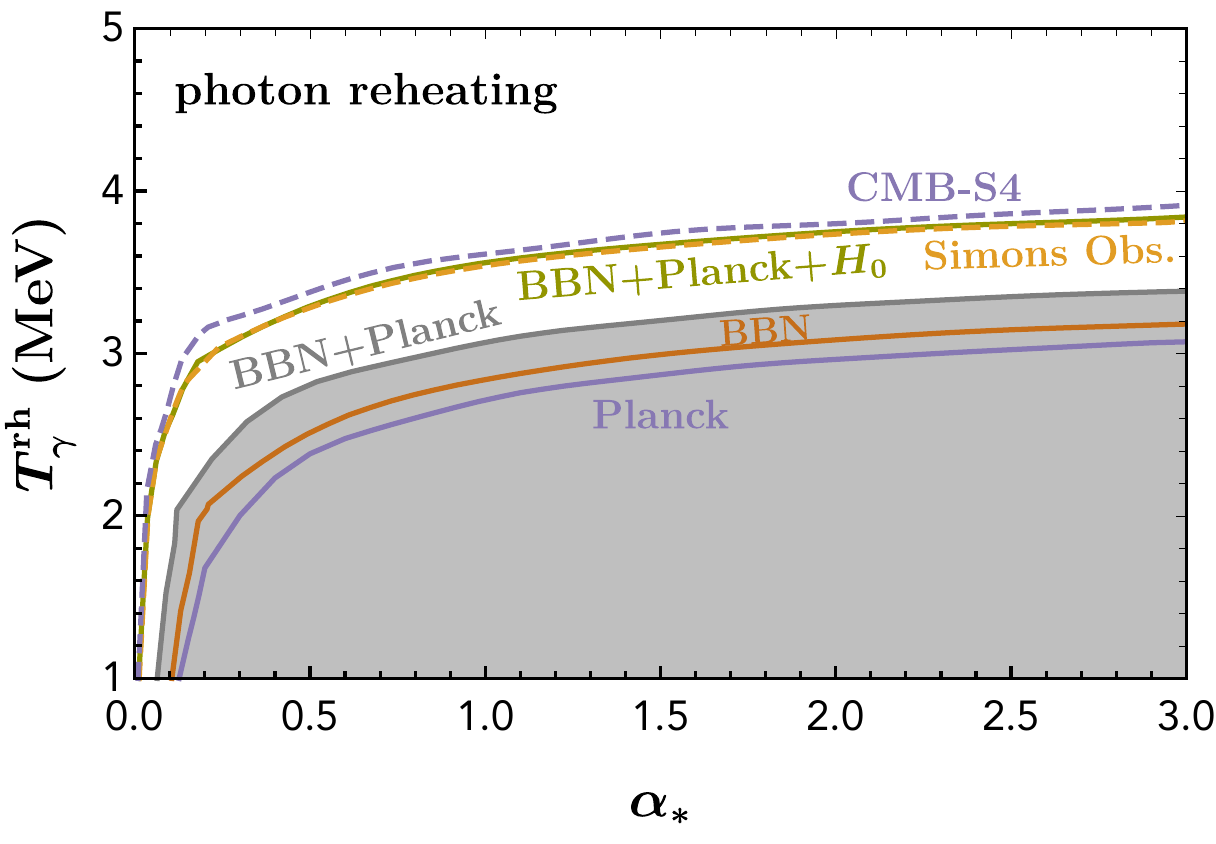} \hspace{3mm}
\includegraphics[width=.48\textwidth]{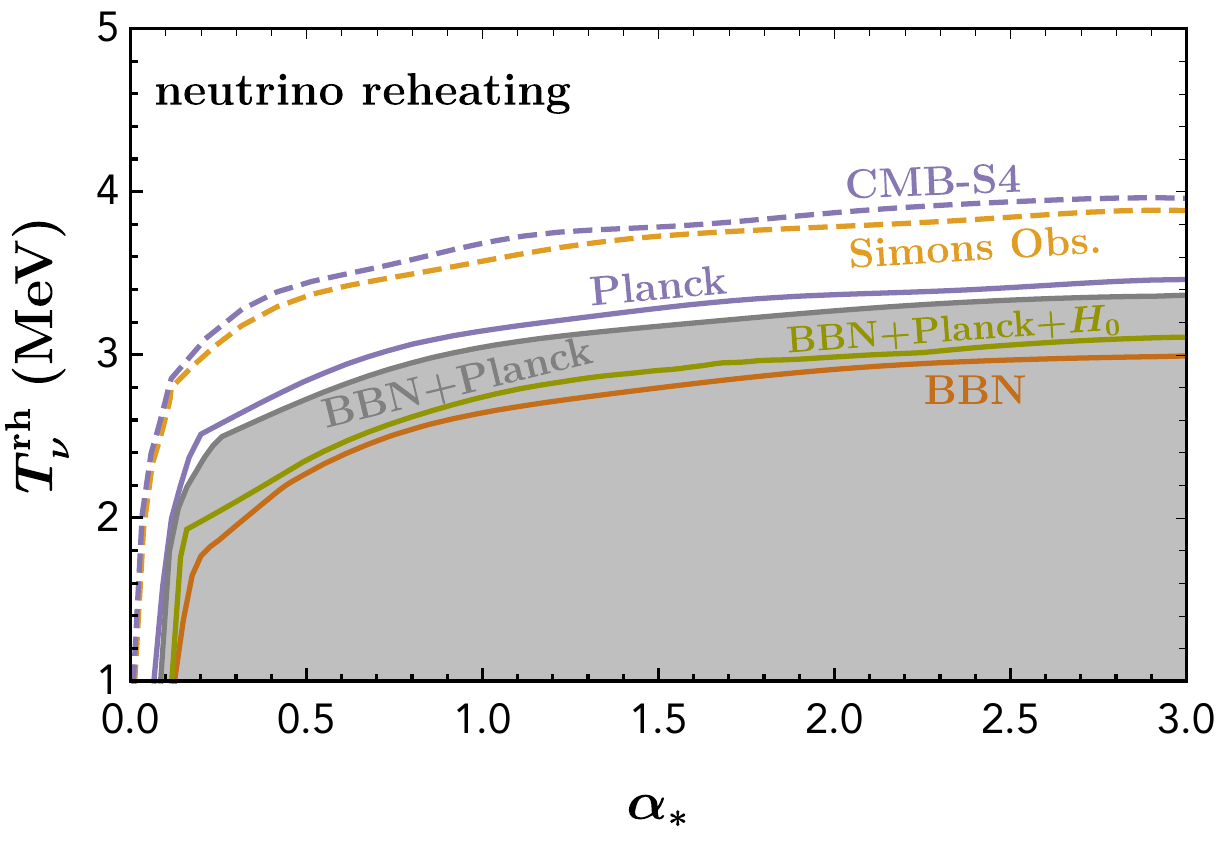}	
	\caption{$95\%$ CL constraints on the phase transition model parameters $\alpha_*$ and $T^{\rm rh}$ from different data sets. The left (right) panel is for the photon (neutrino) reheating case. The $\Omega_b h^2$ are marginalized over for both panels. For a low phase transition temperature $T^{\rm rh} \approx 1$~MeV, the upper bounds on $\alpha_*$ are 0.064 (BBN+Planck), 0.013 (Simons Obs.), 0.010 (CMB-S4) for the photon-reheating case and 0.086 (BBN+Planck), 0.012 (Simons Obs.), 0.010 (CMB-S4) for the neutrino-reheating case. }
	\label{fig:constraint}
\end{figure}

In Fig.~\ref{fig:constraint}, we show the 95\% confidence level (CL) exclusion limits in $\alpha_*$ and $T^{\rm rh}$ after marginalizing over $\Omega_b h^2$. For $\Omega_b h^2 \in [0.0215,0.0230]$, we calculate the minimum of $\chi^2$, $\chi^2_{\rm min}$, and have the 95\% CL limits correspond to $\Delta \chi^{2}= \chi^{2} -\chi^{2}_{\rm min} = 5.99$. For the photon-reheating case, the combination of BBN+Planck is stronger than the individual data set.  Adding the local $H_0$ measurement provides more stringent constraints, such that BBN+Planck+$H_0$ has the most stringent constraints. This is because the local (large) $H_0$ value prefers a larger $N_{\rm eff}$ than the SM one, while the photon-reheating case has a smaller $N_{\rm eff}$. Taking $\alpha_* = 1$ and the BBN+Planck data set, the strong first-order phase transition needs to have $T^{\rm rh}_\gamma > 3$\,MeV. 

For the neutrino-reheating case in the right panel of Fig.~\ref{fig:constraint}, the combination of BBN+Planck provides a weaker constraint than the Planck-only one. This is mainly due to the around two sigma discrepancy of the deuterium abundance from the Planck preferred value for $\Omega_b h^2$. As can be seen from the lower right panel of Fig.~\ref{fig:PandNReh} in Appendix~\ref{sec:plots-observables}, some $T^{\rm rh}_\nu$ for a fixed $\alpha_*$ turns to reduce the  $\chi^2_{\rm BBN}$ and relax the combined constraints. Adding the local $H_0$ measurement, the combination BBN+Planck+$H_0$ provides relatively weak constraints. This is due to the fact that $N_{\rm eff}$ in the neutrino-reheating case can be larger than the SM case, which is preferred by the larger $H_0$ value from local measurement. Taking $\alpha_* = 1$ and the BBN+Planck data set, the strong first-order phase transition has $T^{\rm rh}_\nu > 3$\,MeV. 
 
We also note that the neutrino-reheating case predicts $N_{\rm eff}$ larger than the SM value and hence can reconcile the Hubble tension~\cite{Riess:2019cxk,Aghanim:2018eyx}. The preferred model parameter space has $T^{\rm rh}_\nu \approx 3.2$~MeV and insensitive to $\alpha_*$.

\section{Implications for NANOGrav results}\label{sec:nanograv}

\begin{figure}[tb!]
 	\centering
 	\includegraphics[width=0.6\textwidth]{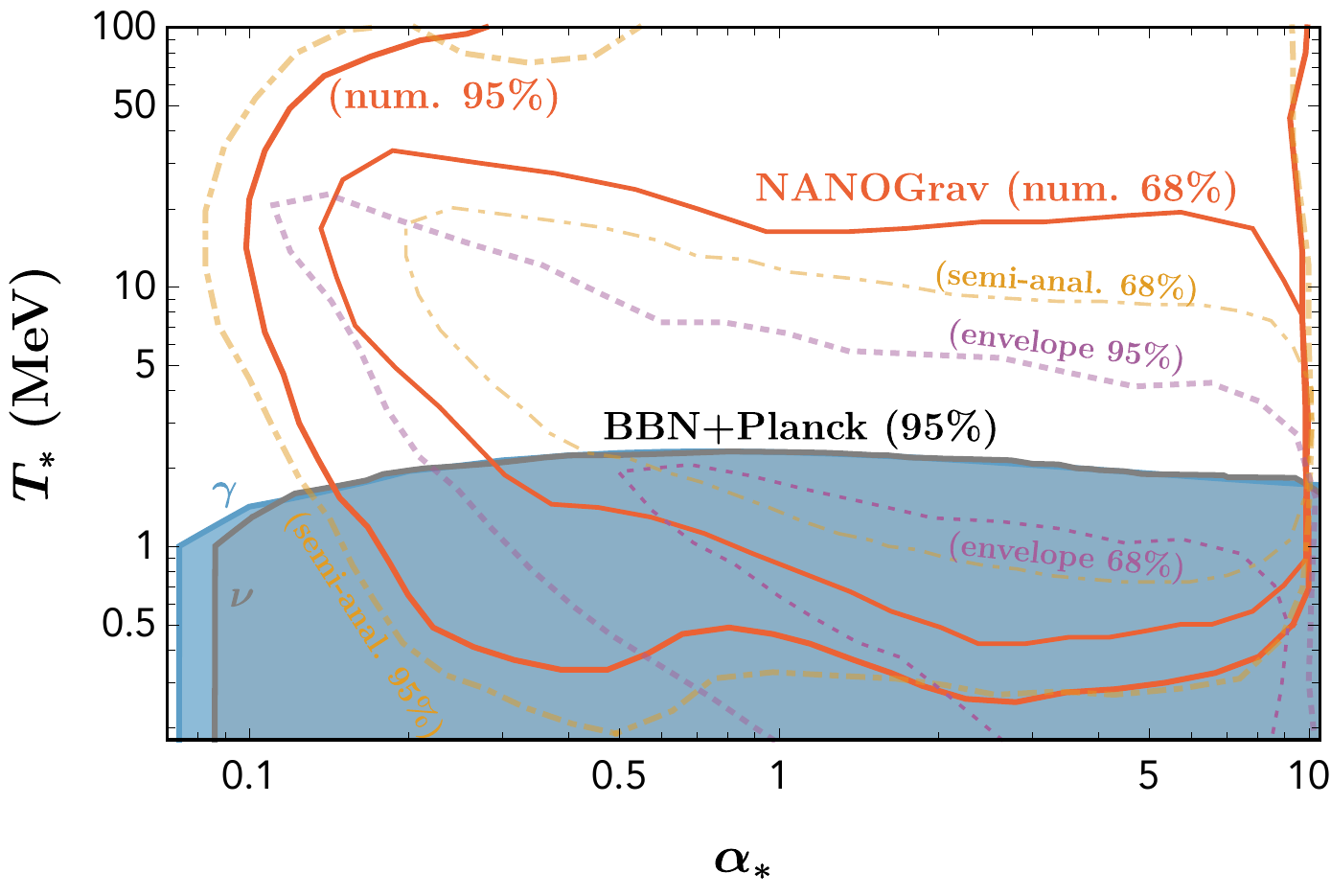}	
  	\caption{The 68\% and 95\% posterior contours in $\alpha_*$ and $T_*$ (equivalent to $T^{\rm p}_{\gamma}$) for the first-order phase transition only explanation for the NANOGrav observation of the stochastic gravitational wave background~\cite{Arzoumanian:2021teu}. The gravitational wave spectra  from numerical, semi-analytic and envelope calculations are shown in the solid, dot-dashed and dotted, respectively. The BBN+Planck constraints at 95\% CL are shown in the shaded regions for both photon- and neutrino-reheating cases. 
 		}
 	\label{fig:NANOGrav_constrint}
 \end{figure}

The SGWB from one pulsar-timing array observatory, NANOGrav, has found a common strain spectrum that could be explained by SMBHBs as well as the common sources from a first-order cosmological phase transition~\cite{Ratzinger:2020koh,Li:2021qer,Bian:2020urb}. The analysis by the NANOGrav collaboration shows that the signal can be explained by a strong first-order phase transition with $\alpha_{*}>0.1$ and a relatively low phase transition temperature below around 20\,MeV~\cite{Arzoumanian:2021teu} (see the 68\% posterior contours in Fig.~\ref{fig:NANOGrav_constrint}). The cosmological constraints on the phase transition strength parameter and temperature can therefore be used to exclude some SGWB signal preferred region.

Using the relations between the temperatures at the percolation time $T^{\rm p}_{\gamma}$ (equivalent to $T_*$ in Ref.~\cite{Arzoumanian:2021teu}) and reheating temperature $T^{\rm rh}_{\gamma}$ in Eqs.~\eqref{eq:phtotonreh}\eqref{eq:neutrinoreh}, the gravitational wave peak frequency $f^0_*$ observed today is related to the peak frequency $f_*$ at the emission time by the scale factor ratio: $f_*^0 = f_*\times a(t_{\rm rh})/a(t_{\rm today})$, where one has assumed $a(t_{\rm rh}) \approx a(t_{\rm p})$. For different phase transition model parameters, the ratio $a(t_{\rm rh})/a(t_{\rm today})$ can be different from the SM case by a factor of order unity that has not been taken into account in Ref.~\cite{Arzoumanian:2021teu} because of its almost negligible effects for the final model inference in Fig.~\ref{fig:NANOGrav_constrint}. The BBN+Planck constraints on $\alpha_*$ and $T_*$ are shown in Fig.~\ref{fig:NANOGrav_constrint} for both photon- and neutrino-reheating cases. The phase transition temperature $T_*$ below around 2 MeV are excluded for an order-one $\alpha_*$. Note that the 68\% region based on the envelope approximation is completed excluded by the cosmological constraints. For the numerical and semi-analytic based spectra, a sizable fraction of SGWB-preferred model parameter space has been constrained by the cosmological data.             
         
\section{Discussion and conclusions}
\label{sec:conclusion}

In this article, we have mainly concentrated on strong first-order phase transitions. For a weaker one with a low phase transition temperature, the cosmological data is also complimentary to the gravitational wave detections to probe phase transition parameter space. In Fig.~\ref{fig:constraint}, the small $\alpha_*$ region for $T^{\rm rh} \leq 1$~MeV is currently constrained at the 0.1 level and can be constrained at the 0.01 level by future CMB experiments. On the other hand, the Square Kilometer Array (SKA) telescope~\cite{Janssen:2014dka} can obtain an improved sensitivity compared to NANOGrav and can also probe $\alpha_*$ at the 0.01 level with around ten years of observation~\cite{Breitbach:2018ddu} (see \cite{Croon:2020cgk} for uncertainties related to gravitational wave spectrum calculations). This complementarity of two different approaches to probe phase transitions will be crucial to distinguish the SMBHB and phase transition explanations to observed gravitational waves.  

In summary, the cosmological constraints on the first-order phase transition strength parameter $\alpha_*$ and temperature $T_*$ (equivalent to $T^{\rm p}_{\gamma}$, the temperature at the bubble percolation time) have been derived based on primordial light element abundances from BBN physics and the effective radiation degrees of freedom $N_{\rm eff}$ for CMB observables. For a strong phase transition with $\alpha_* \sim 1$, the phase transition temperature is constrained to be $T_* \gtrsim 2$~MeV using the BBN+Planck data for both photon- and neutrino-reheating cases. For a weak phase transition with $T_* \leq 1$~MeV and also using the BBN+Planck data, the phase transition strength is constrained to have $\alpha_* <  0.064$ for the photon-reheating case and $\alpha_* <  0.086$ for the neutrino-reheating case.

\subsubsection*{Acknowledgements}
The work is supported by the U.S.~Department of Energy under the contract DE-SC-0017647.

\appendix

\section{Temperature evolution}
\label{sec:Temp_evol}

In this section, we solve for time dependence of photon and neutrino temperatures. We assume that all species involved are thermalized (at least among itself) and are described by either Fermi-Dirac or Bose-Einstein distributions with a corresponding temperature. This is easily justified for the photon-reheating case because the QED interaction keeps the same temperature for photon and electron (positron). While for the neutrino-reheating case, the weak interaction is sufficient to keep neutrinos in thermal equilibrium for most of the boundary of constrained parameter space with a reheating temperature around 3 MeV. For the low-reheating temperature region, we assume that some additional self-interactions among neutrinos (see Appendix~\ref{App:particlemodel}) keep neutrinos in thermal equilibrium even after neutrinos decouple from the photon plasma. Furthermore, we consider that all three-flavor neutrinos are described by a common temperature $T_{\nu}$. This is justified, as for $T_{\nu}\gtrsim 3 \, \text{MeV}$ all species are in thermal equilibrium with photon and thus have a common temperature; while for $T_{\nu}\lesssim 3-5 \, \text{MeV}$ the neutrino oscillation effects become active and all flavors tend towards a common thermal distribution~\cite{Escudero_2019}. 
 
Allowing for energy transfer between photon and neutrino plasmas, the Friedmann equations for photon and neutrino temperatures lead to the following differential equations~\cite{Abenza_2020}:
 \begin{align}\label{eq:tempdiffeq}
\frac{dT_{\nu}}{dt} &= - H\,T_{\nu} + \frac{\delta \rho_{\nu}/\delta t}{\, d\rho_{\nu}/dT_{\nu}} \,\, , \nonumber \\
\frac{dT_{\gamma}}{dt} &= - \frac{4 H \rho_{\gamma}+ 3 H (\rho_{e}+p_{e}) + 3 \,d\rho_{\nu}/dt + 3 H\,T_{\gamma}\,dP_{\text{int}}/dT_{\gamma}}{d\rho_{\gamma}/dT_{\gamma} + d\rho_{e}/dT_{\gamma} + T_{\gamma}\, d^{2}P_{\text{int}}/dT_{\gamma}^{2}} \,\, .
\end{align}
Here, $\rho_{i}$ and $p_{i}$ are the energy density and pressure for the corresponding species. $P_{\rm int}$ and its derivatives account for finite temperature corrections (see Ref.~\cite{Escudero_2019} for details). Here, $\delta \rho_{\nu}/\delta t$ accounts for the energy transfer rate between neutrino and photon plasmas (see Ref.~\cite{Abenza_2020} for formulas). Using the Fermi-Dirac statistics for neutrinos in the rate with $m_{e}=0$, one has a simple formula
\beqa
\frac{\delta \rho_{\nu}}{\delta t} = \frac{G_{F}^{2}}{\pi^{2}}f_{p}\left[32\, f_{a}^{\rm FD}\, (T_{\gamma}^{9}-T_{\nu}^{9})+56\, f_{s}^{\rm FD}\,T_{\gamma}^{4}\,T_{\nu}^{4}\,(T_{\gamma}-T_{\nu})\right]\,\, ,
\eeqa
where $G_{F}$ is the Fermi's constant, $f_{a}^{\rm FD}=0.884$, $f_{s}^{\rm FD}=0.829$, $f_{p}=1.121$. We also take into account the electron mass effects in our numerical calculations as in~\cite{Abenza_2020}. 

\begin{figure}[t!]
	\centering
	\includegraphics[width=0.47\textwidth]{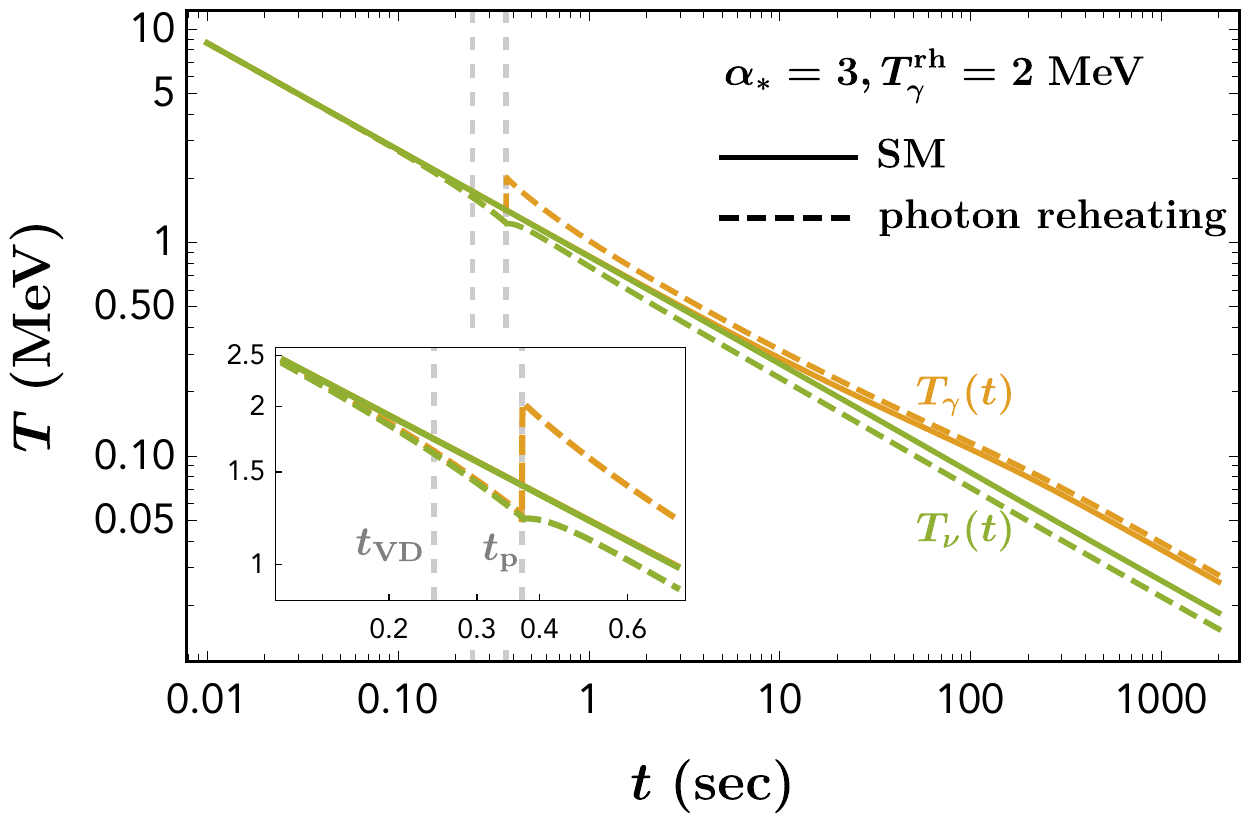}
	\hspace{3mm}
	\includegraphics[width=0.47\textwidth]{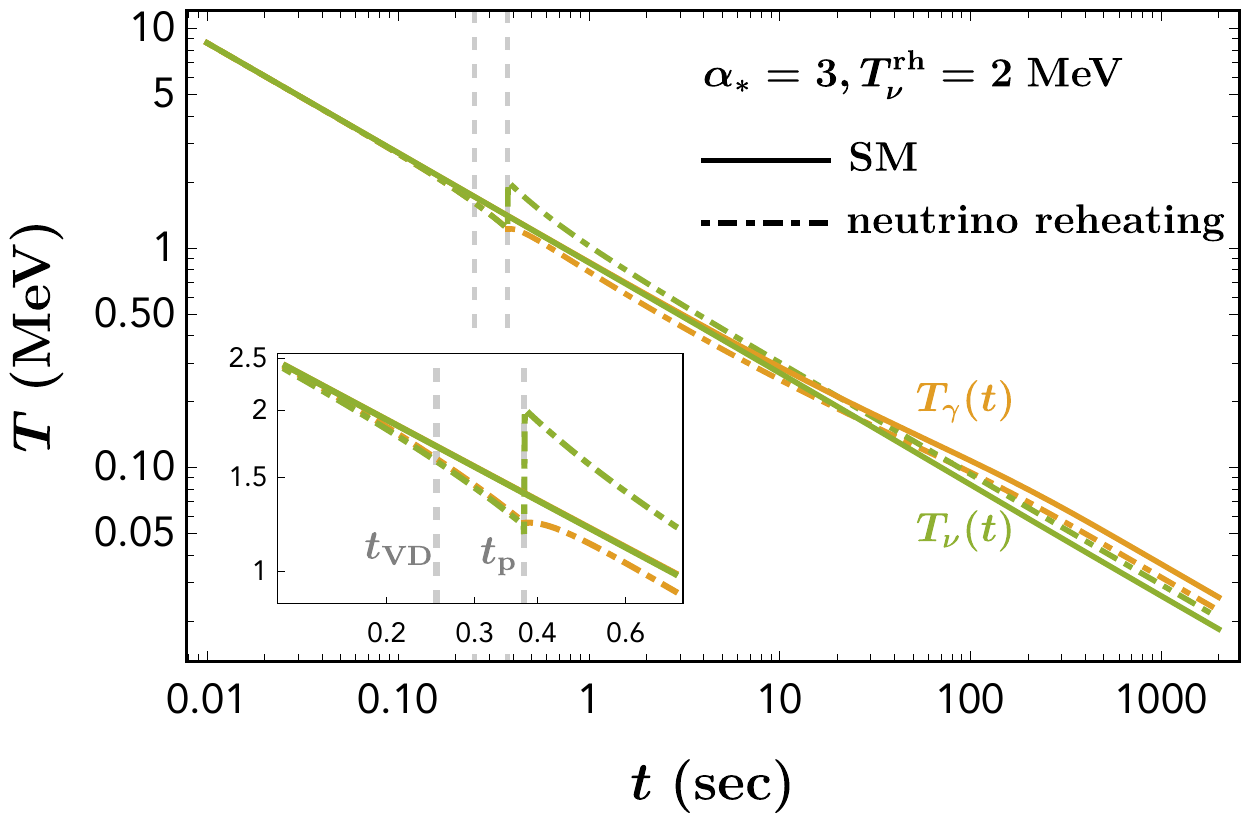}
	\caption{Photon (orange) and neutrino (green) temperatures as a function of time with solid lines showing the SM case.  Left panel: the photon-reheating case with $\alpha_{*}=3$ and $T_{\gamma}^{\rm rh}=2 \,\rm MeV$. The period between the two gray vertical dashed lines has $t_{\rm VD} < t < t_{\rm p}$, representing the supercooling period when the vacuum energy dominates the Hubble rate. Right panel: the neutrino-reheating case with $\alpha_{*}=3$ and $T_{\nu}^{\rm rh}=2 \,\rm MeV$.  
		}\label{fig:Temp-evolution}
\end{figure}

We evaluate the temperatures starting from a relatively high temperature $T_{\gamma}=T_{\nu}=10 \, \text{MeV}$, where the photon and neutrinos are in a common thermal plasma. At this temperature before the phase transition, the corresponding time is  $t=1/(2 H_{\rm bPT})$ with $H_{\rm bPT}=\sqrt{8\pi(\sum_{i}\rho_{i}+\rho_{h})/3 M_{\rm pl}^{2}}$ with $M_{\rm pl}=1.22 \times 10^{19} \, \rm GeV$,  $\rho_i$ representing visible-sector energy densities and $\rho_{h}=\Delta V$ as the hidden sector energy density which is dominated by the vacuum energy. After the phase transition, we consider instantaneous reheating with the vacuum energy transferred to either photon (including $e^\pm$) or neutrinos and to increase their temperatures according to \eqref{eq:phtotonreh} or \eqref{eq:neutrinoreh}. The reheated temperatures become the initial conditions for their later evolution with the Hubble rate after the phase transition given by $H_{\rm aPT}=\sqrt{8\pi(\sum_{i}\rho_{i})/3 M_{\rm pl}^{2}}$ (assuming no additional radiation energy in the hidden sector). In Fig.~\ref{fig:Temp-evolution}, we show the photon and neutrino temperature evolutions as a function of time. For comparison, we also show the SM case, where the photon temperature differs from the neutrino temperature because of entropy transfer after electron-positron annihilation leading to a final temperature ratio of $T_{\gamma}(t_{\rm today})/T_{\nu}(t_{\rm today}) \approx 1.396$~\cite{Abenza_2020}. For the photon-reheating case, the photon temperature has the first increasing moment at the reheating time that is very close to the percolation time $t_{\rm p}$ and the second increasing moment when electron-positron annihilations happen. There is a short period of supercooling where the vacuum energy dominates the Hubble expansion rate, as shown by the dashed lines in Fig.~\ref{fig:Temp-evolution} and magnified in the inset plot. During the supercooling period, the temperature drops exponentially in time from $T_{\rm VD} = \left(30\,\Delta V/\pi^{2} g_{*,\rm SM}\right)^{1/4}$ (at time $t_{\rm VD}$) with $g_{*, \rm SM}\approx 10.75$, corresponding to the start of vacuum energy dominance (VD), to $T_{\gamma}^{\rm p}$ at the percolation time. Note that the final neutrino temperature is lower than the SM case because the supercooling effect dominants the energy transferred from the heated-up photon sector to the neutrino sector. 

For the neutrino-reheating case shown in the right panel of Fig.~\ref{fig:Temp-evolution}, neutrinos get heated up first when the hidden sector dumps energy into them. If this happens earlier before neutrinos thermally decouple from the photon plasma, neutrinos transfer energy into the photon plasma, which slightly increases the photon temperature. After the electron and positron annihilations, the photon temperature increases slightly again. As a net result, the ratio of the final photon temperature over the neutrino temperature is smaller than the SM case. Note that the asymptotic ratio of $T_{\gamma}/T_{\nu}$ is close to one in this right panel. This is rather accidental and depends on the specific choices of parameters here.

\begin{figure}[tb!]
	\centering
	\includegraphics[width=0.48\textwidth]{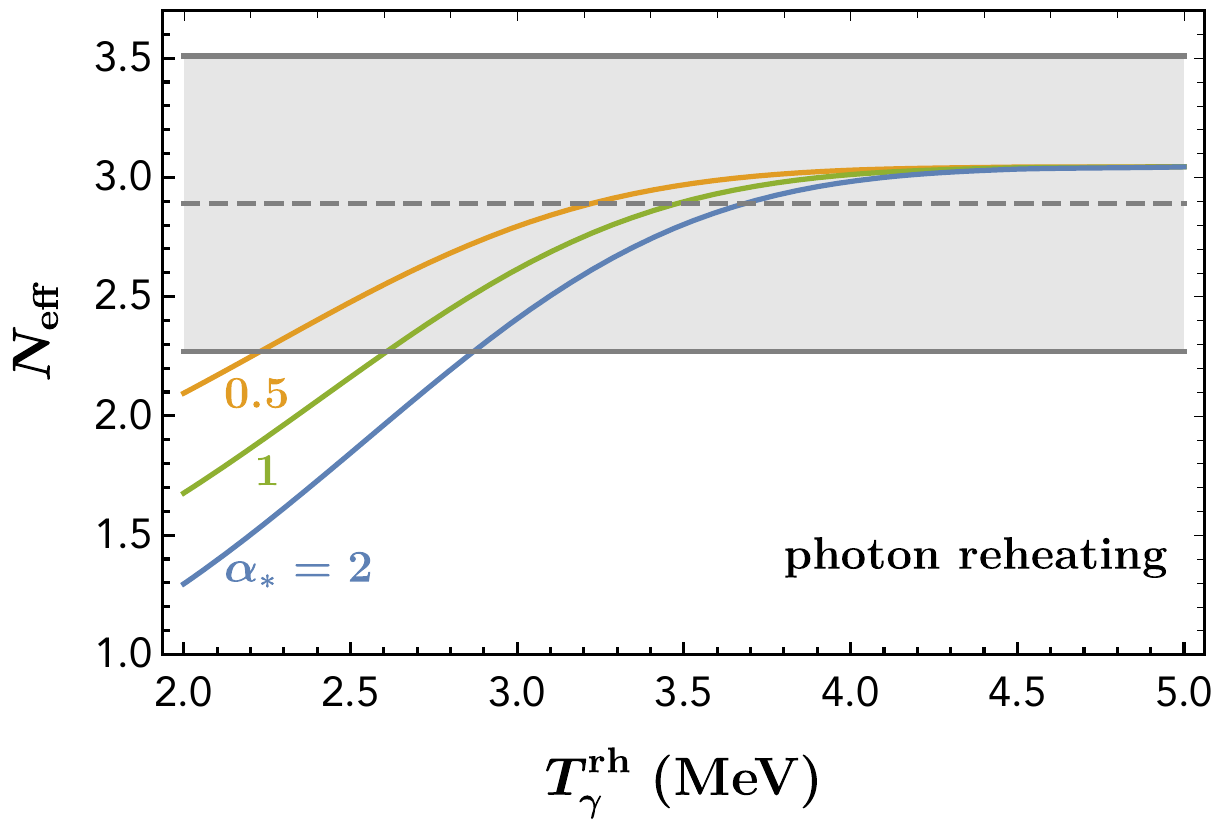}
	\hspace{3mm}
	\includegraphics[width=0.47\textwidth]{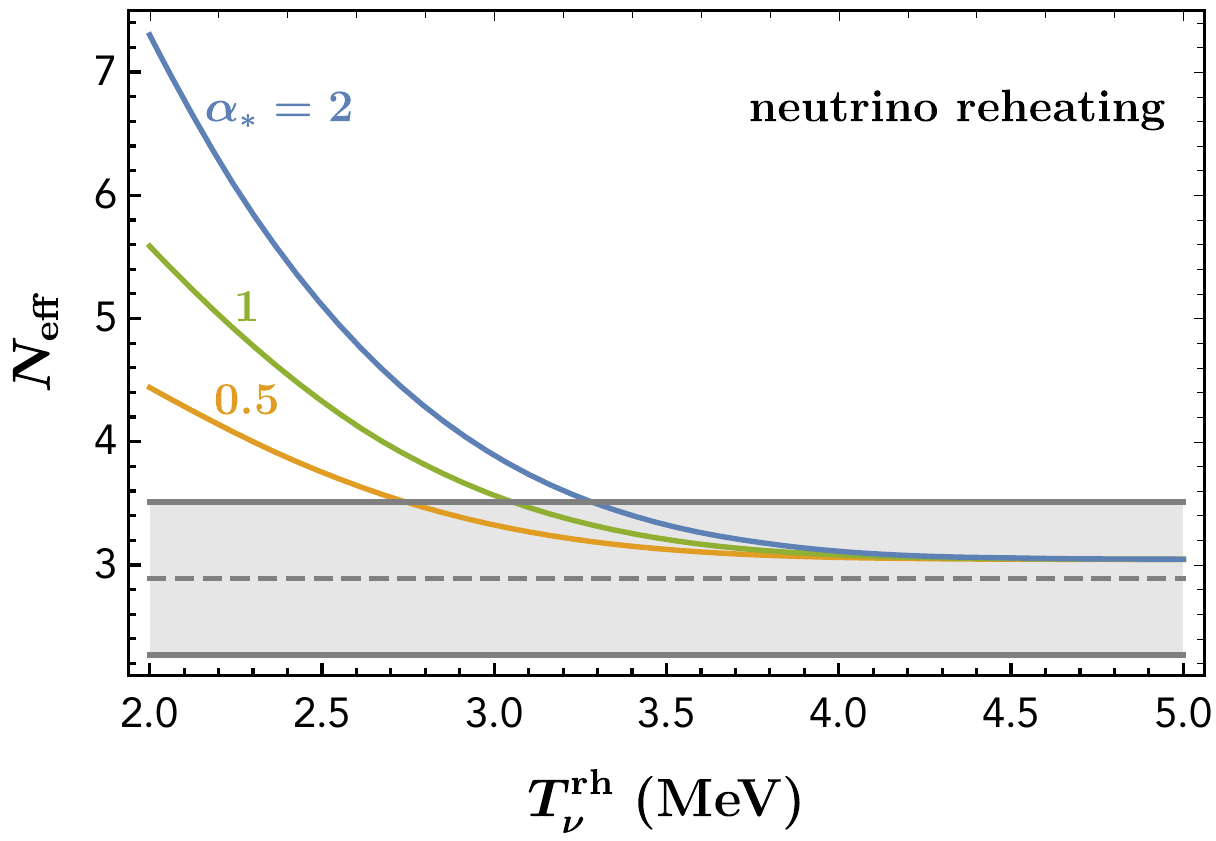}
	\caption{Left panel: $N_{\rm eff}$ for different photon-reheating temperatures $T_{\gamma}^{\rm rh}$ and $\alpha_*$. The gray shaded region represents the $\pm 2\sigma$ range from the CMB data by the Planck collaboration~\cite{Aghanim:2018eyx}. Right panel: the same as the left one, but for the neutrino-reheating case.	}
	\label{fig:PandNReh_Neff}
\end{figure}

\section{{\boldmath{$N_{\rm eff}$}}, {\boldmath{$Y_{\rm P}$}}, {\boldmath{$\mbox{D}/\mbox{H}|_{\rm P}$}} for different {\boldmath $T^{\rm rh}$} and {\boldmath $\alpha_*$}}
\label{sec:plots-observables}

For the SM case, we have checked our numerical calculations to reproduce the SM value of $N_{\rm eff}=3.045$~\cite{Abenza_2020}. For the phase transition cases, we show $N_{\rm eff}$ for different phase transition temperatures and $\alpha_*$ in Fig.~\ref{fig:PandNReh_Neff}. For the photon-reheating case in the left panel, one can see that a smaller $T_{\gamma}^{\rm rh}$ or a larger $\alpha_*$ corresponds to a smaller $N_{\rm eff}$ compared to the SM case. This is manifest from the left panel of Fig.~\ref{fig:Temp-evolution}, where the ratio of the final neutrino over the photon temperature is reduced. For the neutrino-reheating case in the right panel, on the other hand, $N_{\rm eff}$ increases as $T_{\nu}^{\rm rh}$ decreases or $\alpha_*$ increases. Again, this is anticipated as the related temperature profiles are modified compared to the SM case, shown in the right panel of Fig.~\ref{fig:Temp-evolution}. 

\begin{figure}[tb!]
	\centering
       \includegraphics[width=0.48\textwidth]{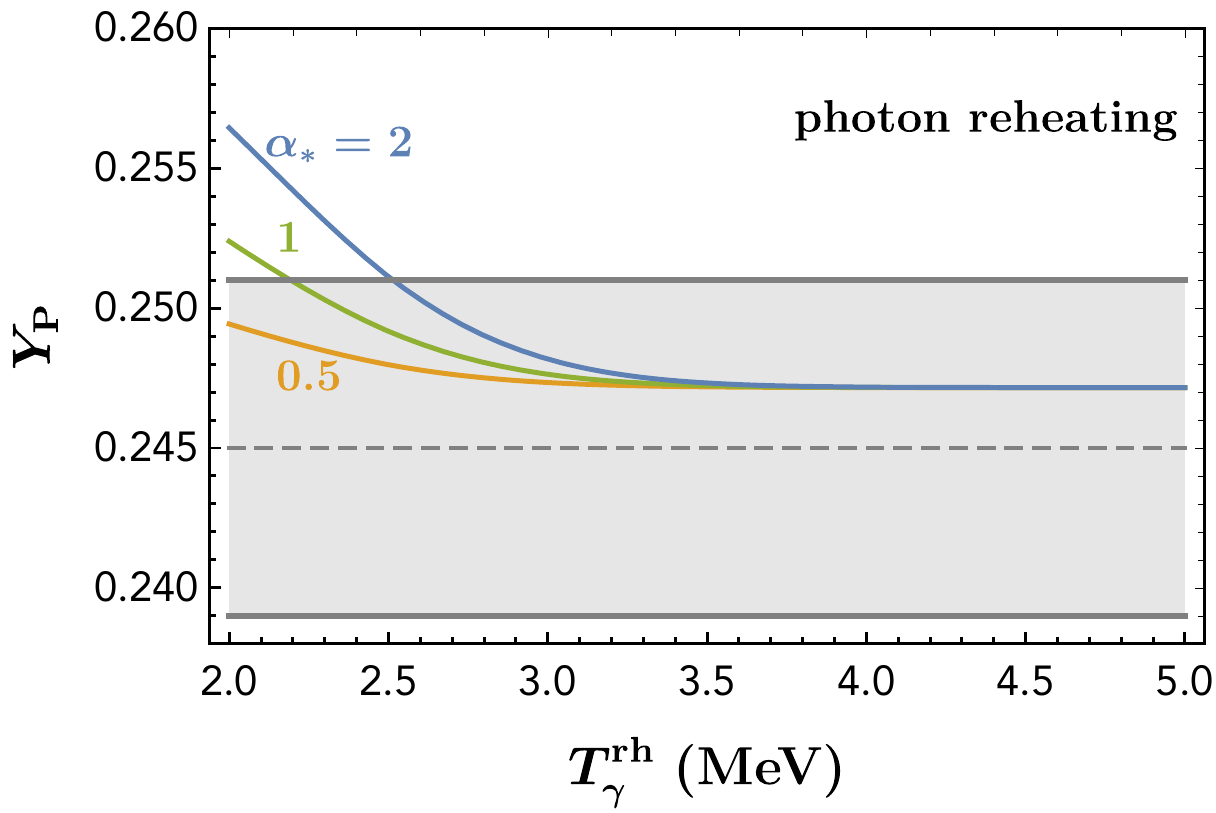}       \hspace{3 mm}
       \includegraphics[width=0.48\textwidth]{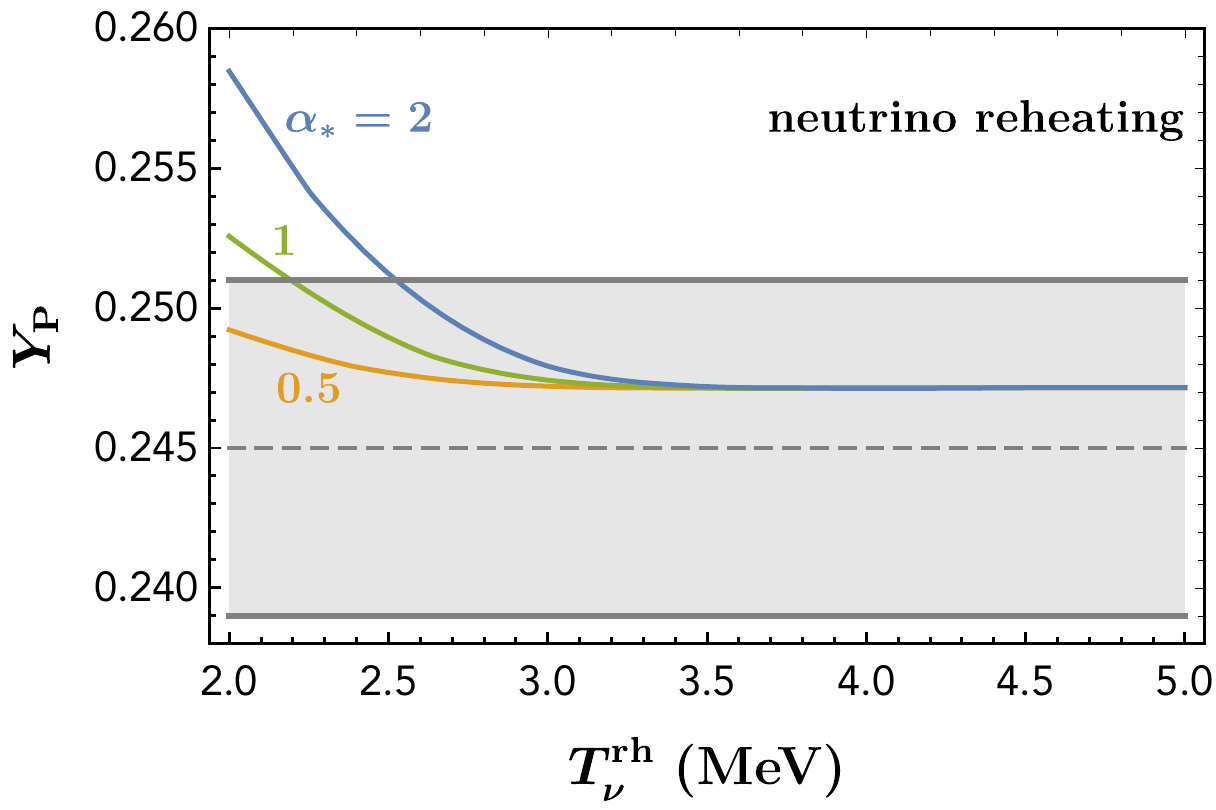} \\
        \includegraphics[width=0.46 \textwidth]{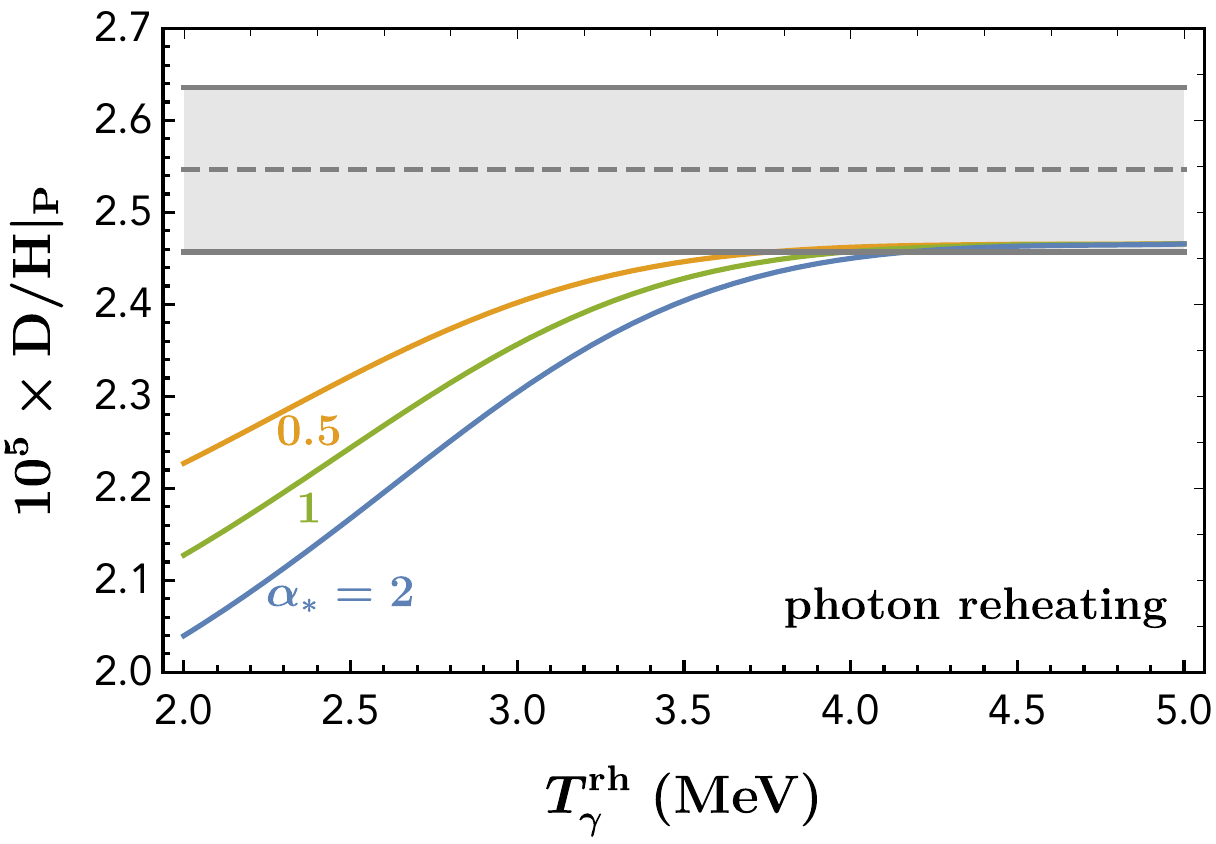}    \hspace{5 mm}
       	\includegraphics[width=0.46 \textwidth]{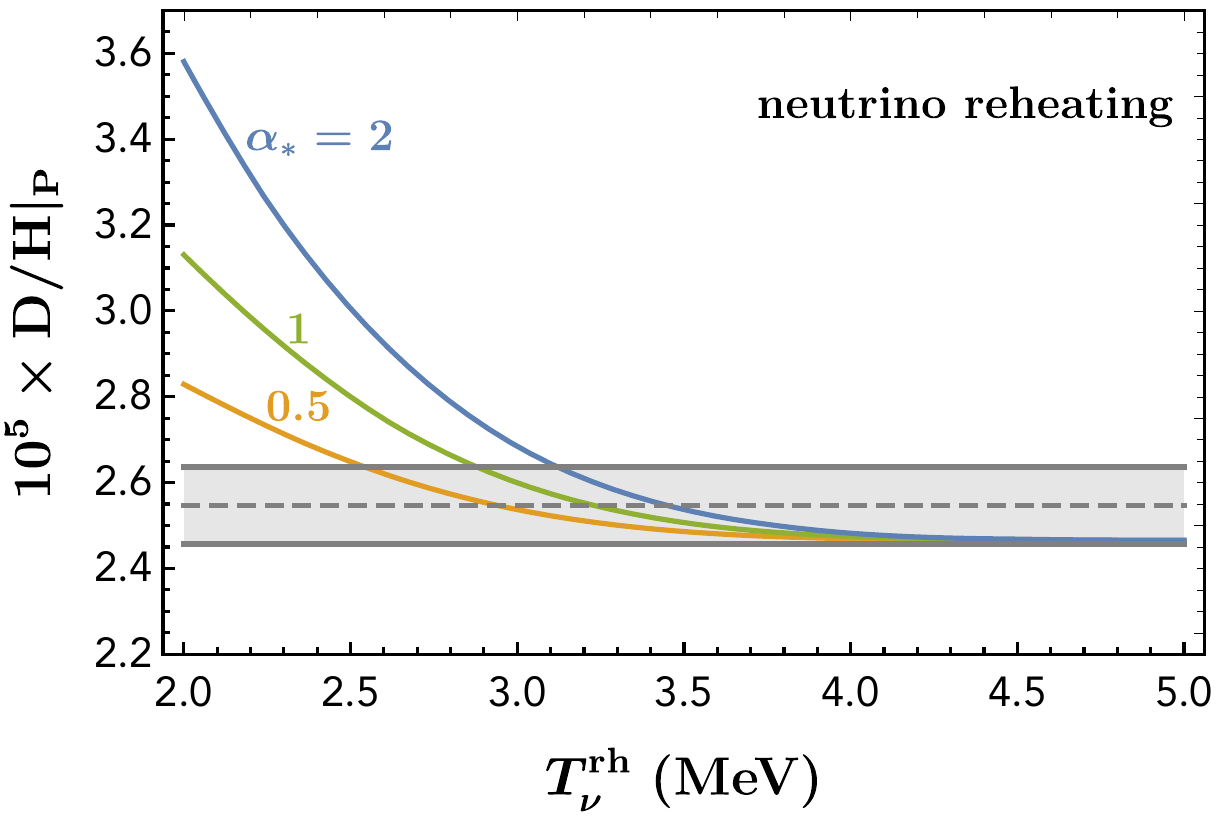}
 \caption{Upper panels: the helium abundance ratio $Y_{\rm P}$ as a function of reheating temperature for different $\alpha_*$ and fixed $\Omega_{b}h^{2}=0.0223$ and $\tau_{n}=879.4$~sec. The gray shaded band shows the $\pm 2\sigma$ region of the measured values~\cite{ParticleDataGroup:2020ssz,Pitrou:2020etk}. Lower panels: the same as the upper ones, but for the deuterium abundance ratio.  
 }
 \label{fig:PandNReh}
 \end{figure}

In the upper two panels of Fig.~\ref{fig:PandNReh}, we show $Y_{\rm p}$ as a function of reheating temperatures for different values of $\alpha_*$. Also shown in the gray band is the $\pm 2 \sigma$ range around the measured value.  In the lower two panels of Fig.~\ref{fig:PandNReh}, we show the deuterium abundance ratios for different model parameters. The gray band shows the $\pm 2 \sigma$ range around the measured value~\cite{ParticleDataGroup:2020ssz,Pitrou:2020etk}. Comparing the upper and lower panels, one can see that the deuterium abundance is likely to provide the dominant BBN constraint. Also, note that the deuterium abundance is approximately proportional to $N_{\rm eff}$ (see Fig.~\ref{fig:PandNReh_Neff} and the lower panels of Fig.~\ref{fig:PandNReh}). On the other hand, the helium abundance ratio (the upper panels of Fig.~\ref{fig:PandNReh}) does not have such a correlation with $N_{\rm eff}$ and hence provides additional constraints from $N_{\rm eff}$.

\section{Detailed BBN and CMB experimental observables}\label{App:dataset}

Here we list the theoretically predicted and observed values for some BBN and CMB observables that are used in Section~\ref{sec:constraint}. For the two BBN observables, we take the central values and the experimental errors from PDG~\cite{ParticleDataGroup:2020ssz} and the theoretical errors from Ref.~\cite{Pitrou:2020etk}
\beqa
\label{eq:BBN-data}
\begin{array}{lll}
 Y_{\rm P}^{\rm obs} = 0.245 \,, &\sigma(Y_{\rm P}^{\rm obs}) = 0.003 \,,  &\sigma(Y_{\rm P}^{\rm theo}) = 0.00014 ~, \\
\rm D/ \rm H|_{\rm P}^{\rm obs}=2.547 \times 10^{-5}\,, &\sigma(\rm D/\rm H|_{\rm P}^{\text{obs}})= 0.025 \times 10^{-5} \,, &\sigma(\rm D/\rm H|_{\rm P}^{\text{theo}}) =  0.037 \times 10^{-5} ~.
\end{array}
\eeqa
The $\chi^2$ is given by 
\beqa
\label{eq:chi2BBN}
    \chi^{2}_{\rm BBN}= \frac{\left[Y_{\rm P}(\Omega_bh^2, \alpha_*, T^{\rm rh}_{\gamma, \nu}) - Y_{\rm P}^{\text{obs}}\right]^{2}}{ \sigma(Y_{\rm P}^{\text{theo}})^{2}+\sigma(Y_{\rm P}^{\text{obs}})^{2}} + \frac{\left[\rm D/\rm H|_{\rm P}(\Omega_bh^2, \alpha_*, T^{\rm rh}_{\gamma, \nu}) - \rm D/\rm H|_{\rm P}^{\text{obs}}\right]^{2}}{\sigma(\rm D/\rm H|_{\rm P}^{\text{theo}})^{2}+\sigma(\rm D/\rm H|_{\rm P}^{\text{obs}})^{2}} \, .
\eeqa

For the CMB data, we take the mean values and covariance matrices deduced in Ref.~\cite{Sabti_2020}. For the Planck-only data, the baseline TTTEEE+lowE analysis has been used for $\Theta \equiv (\Omega_{b}h^{2}, N_{\text{eff}}, Y_{\rm P})$ and has the summed $\chi^2$ given by 
 \beqa
 \label{eq:chi2CMB}
    \chi^{2}_{\text{CMB}} = (\Theta-\Theta_{\rm obs})^{\rm T} \,\Sigma_{\text{CMB}}^{-1}\, (\Theta-\Theta_{\rm obs}) \quad ,\quad \mbox{with}\quad\Sigma_{\text{CMB}} = \begin{bmatrix}
    \sigma_{1}^{2} & \sigma_{1}\sigma_{2}\rho_{12} & \sigma_{1}\sigma_{3}\rho_{13} \\
    \sigma_{1}\sigma_{2}\rho_{12} & \sigma_{2}^{2} & \sigma_{2}\sigma_{3}\rho_{23} \\
    \sigma_{1}\sigma_{3}\rho_{13} & \sigma_{2}\sigma_{3}\rho_{23} & \sigma_{3}^{2}
    \end{bmatrix} ~,
 \eeqa
with the numerical values as
\beqa
\label{eq:planck-only}
\Theta_{\rm obs} &=& (0.02225, 2.89, 0.246)~, \nonumber \\
(\sigma_{1},\sigma_{2},\sigma_{3}) &=& (0.00022,0.31,0.018)~, \\
(\rho_{12},\rho_{13},\rho_{23}) &=& (0.4,0.18,-0.69)~. \nonumber 
\eeqa

For the Planck+BAO+$H_0$ data set, one has 
\beqa
\label{eq:planck-H0}
\Theta_{\rm obs} &=& (0.02345, 3.36, 0.249)~, \nonumber  \\ 
(\sigma_{1},\sigma_{2},\sigma_{3}) &=& (0.00025,0.25,0.020)~,\\
(\rho_{12},\rho_{13},\rho_{23}) &=& (0.011,0.50,-0.64)~. \nonumber 
\eeqa

For future cosmological data from the Simons Observatory~\cite{SimonsObservatory:2018koc}, we take 
\beqa
\label{eq:Simons}
\Theta_{\rm Fiducial} &=& (0.02236, 3.0453, 0.2471)~, \nonumber \\
(\sigma_{1},\sigma_{2},\sigma_{3}) &=& (0.000073, 0.11, 0.0066)~,\\
(\rho_{12},\rho_{13},\rho_{23}) &=& (0.072,0.33,-0.86)~.  \nonumber 
\eeqa
While for CMB-S4~\cite{Abazajian:2019eic}, one has 
\beqa
\label{eq:CMBS4}
\Theta_{\rm Fiducial} &=& (0.02236, 3.0453, 0.2471)~, \nonumber \\
(\sigma_{1},\sigma_{2},\sigma_{3}) &=& (0.000047, 0.081, 0.0043)~,\\
(\rho_{12},\rho_{13},\rho_{23}) &=& (0.25,0.22,-0.84)~.  \nonumber 
\eeqa

\section{Interactions for photon and neutrino reheating}\label{App:particlemodel}

Here, we provide detailed interactions between the phase transition hidden sector and the SM sector such that our instantaneous reheating approximation can be justified. For the delayed reheating case, the constraints on the phase transition parameters are in general more stringent. Ignoring the detailed dynamics to provide a first-order phase transition, we simply use a real scalar field $\Phi$ to represent the order parameter, which has a nonzero vacuum expectation value (VEV), $f \equiv \langle \Phi \rangle$, in the final true vacuum state. The finite-temperature potential $V(\Phi, T)$ provides the first-order phase transition from the high-temperature phase with the false vacuum to the low-temperature phase with the true vacuum. 

For the photon-reheating case, the following dimension-five operator 
\beqa
\label{eq:photon-operator}
\mathcal{O}_5^{\gamma}=\frac{\alpha}{4\pi\,\Lambda}\,\Phi \, F^{\mu\nu}F_{\mu\nu} 
\eeqa
could have the $\phi \equiv \Phi - \langle \Phi \rangle$ particle decay into two photons and transfer the energy from the hidden sector to the visible sector. For the mass $m_\phi \sim 1$~MeV, one stringent constraint comes from the muon $g-2$. The two-loop contributions for the light-by-light and vacuum polarization diagrams have been calculated in Refs.~\cite{Blokland:2001pb,Marciano:2016yhf}, which provides a weak constraint on the cutoff scale $\Lambda \gtrsim 3$~GeV. We also note that to UV-complete the operator in \eqref{eq:photon-operator}, additional electrically-charged particles are generally needed and have a mass above $\mathcal{O}(100\,\mbox{GeV})$ to evade collider constraints. Therefore, the cutoff scale is generically $\Lambda \gtrsim 100$~GeV. Requiring its decay width $\Gamma_\phi = \alpha^2\,m_\phi^3/(64\pi^3\Lambda^2)$ much larger than the Hubble rate at the reheating temperature, $H_{*} \approx 4 \times 10^{-25} \, \mbox{GeV}\times [g_{*}(t_{\rm rh})/10.75]^{1/2} (T_{\gamma}^{\rm rh}/1 \, \rm MeV)^{2}$, the cutoff scale has an upper bound as $\Lambda \lesssim 8$~TeV for $m_\phi \sim 1$~MeV, which easily satisfies the existing constraints. 

At the same dimension-five level, one can have the following operator for $\Phi$ to decay into a pair of electron and positron, which are tightly coupled to the photon plasma
\beqa
\mathcal{O}_5^{e}=\frac{\Phi\, H\,\overline{L}_{L}e_{R}}{\Lambda} ~,
\eeqa
with $L_{L}$ as the left-handed weak doublet and $e_{R}$ as the right-handed electron field. For the couplings to muon and tau leptons, additional flavor structures beyond the SM can appear. Here, we simply assume that the flavor matrix is proportional to the SM Higgs Yukawa matrix of charged leptons. After electroweak symmetry breaking and $\Phi$ developing a VEV, one has a Yukawa coupling of $\phi$ to electron as $(v/\Lambda)\,\phi\,\overline{e} e$ with $v \approx 174$~GeV as the electroweak scale. Requiring the decay width $\Gamma_\phi = v^2\,m_\phi/(8\pi\,\Lambda^2)$ (for $m_\phi > 2 m_e$) larger than $H_*$ at the MeV temperature, one needs to have $\Lambda < 2\times 10^{12}$~GeV, which is allowed by the constraint $\Lambda \gtrsim 10^7 \times v \approx 10^9$~GeV from the MeV-range scalar coupling to electrons~\cite{Liu_2016}. 

For the neutrino reheating case, one could introduce the following dimension-six operator 
\beqa
\mathcal{O}_6^{\nu} = \frac{\Phi (H L_L)^2}{\Lambda^2} 
\eeqa
to have $\phi$ decay into two neutrinos after the electroweak symmetry breaking. Similar to previous two cases, one can have the decay width much larger than the Hubble scale at the MeV temperature while satisfying laboratory constraints with $\Lambda \gtrsim 1\times 10^4$~GeV~\cite{Blum:2018ljv,Brune:2018sab,Blinov_2019}. Also note that the MeV-scale $\phi$ particle can contribute to neutrino self-interactions and keep neutrinos in thermal equilibrium till a later time. For $T_\nu \sim m_\phi$, the neutrino self-interaction rate is $\Gamma_{\nu\nu} =  n_{\nu} \langle\sigma v\rangle_{\nu \nu} \sim v^8\,T_\nu/\Lambda^8$, which is faster than the Hubble rate for $\Lambda \lesssim 1\times10^6$~GeV.


\providecommand{\href}[2]{#2}\begingroup\raggedright\endgroup

\end{document}